\documentclass{amsart}
\date{\today}
\usepackage{amssymb}
\usepackage{latexsym}
\newcommand{\Z}{{\mathbb Z}}
\newcommand{\R}{{\mathbb R}}

\newcommand{\N}{{\mathbb N}}

\newcommand{\beq}{\begin{eqnarray}}
\newcommand{\eeq}{\end{eqnarray}}
\def\ord{{\rm ord\,}}
\def\tr{{\rm tr\,}}

\newtheorem{theorem}{Theorem}

\newtheorem{lemma}{Lemma}[section]
\newtheorem{prop}[lemma]{Proposition}
\newtheorem{coro}[lemma]{Corollary}

\begin{document}
\title[Quantum Dynamics in One Dimension II.]{Power-law Bounds on Transfer Matrices
and Quantum Dynamics in One Dimension II.}
\author[D.\ Damanik, A.\ S\"ut\H{o}, S.\ Tcheremchantsev]{David Damanik$\, ^1$,
Andr\'as S\"ut\H{o}$\, ^2$, and Serguei Tcheremchantsev$\, ^3$}
\thanks{D.\ D.\ was supported in part by NSF grant DMS--0227289\\
\hspace*{9.5pt} A.\ S.\ acknowledges support by OTKA through
grants T 030543 and T 042914} \maketitle \vspace{0.3cm} \noindent
$^1$ Department of Mathematics 253--37, California Institute of
Technology, Pasadena,
CA 91125, USA, E-mail: \mbox{damanik@its.caltech.edu}\\[2mm]
$^2$ Research Institute for Solid State Physics, Hungarian Academy of Sciences,
 P.O.B.~49, H-1525~Budapest, Hungary, E-mail: \mbox{suto@szfki.hu}\\[2mm]
$^3$ UMR 6628 -- MAPMO, Universit\'{e} d'Orleans, B.P.~6759, F-45067 Orleans
C\'{e}dex, France, E-mail: \mbox{serguei.tcherem@labomath.univ-orleans.fr}\\[3mm]

\begin{abstract}
We establish quantum dynamical lower bounds for a number of
discrete one-dimensional Schr\"odinger operators. These dynamical
bounds are derived from power-law upper bounds on the norms of
transfer matrices. We develop further the approach from part~I and
study many examples. Particular focus is put on models with
finitely or at most countably many exceptional energies for which
one can prove power-law bounds on transfer matrices. The models
discussed in this paper include substitution models, Sturmian
models, a hierarchical model, the prime model, and a class of
moderately sparse potentials.
\end{abstract}

%
%
%
%

\section{Introduction}

Consider a discrete one-dimensional Schr\"odinger operator
\begin{equation}\label{oper}
H \psi(n)=\psi(n-1)+\psi(n+1) + V(n)\psi(n),
\end{equation}
in $\ell^2(\Z)$ or $\ell^2(\N)$ (with a Dirichlet boundary
condition). We are interested in proving lower bounds on the
spreading of an initially localized wavepacket under the dynamics
governed by $H$. That is, if we consider the initial state $\psi$,
we ask how fast $\psi(t)=\exp (-itH) \psi$ spreads out. One is
normally interested in initial states that are well localized. In
the present paper we shall limit our attention to the case $\psi =
\delta_1$. 

A typical quantity that is considered to measure the spreading of
$\psi(t)$ is the following: Define
\begin{equation}\label{mpo}
\langle |X|_\psi^p \rangle (T) = \sum_n |n|^p a(n,T),
\end{equation}
where
\begin{equation}\label{ant}
a(n,T)=\frac{1}{T} \int_0^{+\infty} e^{-2t/T} | \langle \delta_n ,
\psi(t) \rangle |^2 \, dt.
\end{equation}
Clearly, the faster $\langle |X|_\psi^p \rangle (T)$ grows, the
faster $\psi(t)$ spreads out, at least averaged in time. One
typically wants to prove power-law lower bounds on $\langle
|X|_\psi^p \rangle (T)$ and hence it is natural to define the
following quantity: For $p > 0$, define the lower growth exponent
$\beta^-_{\psi}(p)$ by
$$
\beta^-_{\psi}(p)=\liminf_{T \to +\infty} \frac{{\rm log} \,
\langle |X|_\psi^p \rangle (T) }{ {\rm log} \, T}.
$$

There are presently two distinct approaches to proving lower
bounds for $\beta^-_{\psi}(p)$. The first goes back to works of
Guarneri \cite{g}, Combes \cite{c}, and Last \cite{l} and is based
on a study of the Hausdorff dimension of the spectral measure
$\mu_\psi$ associated with the pair $(H,\psi)$. Namely, we have
the following bound:
\begin{equation}\label{in4b}
\beta^-_{\psi} (p) \ge p \cdot {\rm dim}_H(\mu_\psi).
\end{equation}
The Jitomirskaya-Last extension \cite{jl1,jl2} of Gilbert-Pearson
theory \cite{gp} allows for a convenient way of investigating
${\rm dim}_H(\mu_\psi)$ and hence this approach has enjoyed some
popularity (see, e.g., \cite{dkl,kls2,z} for applications).

On the other hand, this bound clearly gives nothing in the case of
a zero-dimensional spectral measure, for example, in the case of a
pure point measure. There are a number of models where one expects
(or can prove) pure point spectrum with strictly positive values
for $\beta^-_{\psi}(p)$. An example is given by the random dimer
model; studied, for example, in \cite{bg,DWP,jss}. It is therefore
desirable to have a way of proving lower bounds on the transport
exponents which works for such models and, of course, whose input
is easy to verify in concrete cases. Such an approach was
developed in \cite{dt} (and employed in \cite{jss} to prove the
conjectured dynamical lower bound for the random dimer model), and
the present article is a continuation of that paper. The necessary
input are power-law upper bounds on transfer matrices for certain
energies. It may come as a surprise that dynamical bounds can be
obtained if there is only one energy where one can exhibit a
power-law bound for the transfer matrix. This is indeed necessary
for models such as the random dimer model and related ones
\cite{dss}, where there are only a finite number of such energies.

Another advantage of the approach from \cite{dt} over the bound
\eqref{in4b} is the stability of its input with respect to
perturbations of the potential $V$. It was noted in \cite{dt} that
if its approach can be applied to a given model, then it can also
be applied to all finitely supported perturbations of the given
potential---and it gives the same dynamical bounds for the
perturbed models. Such a stability is not true in general for
bounds derived using \eqref{in4b}. For example, it may happen that
the addition of a finitely supported perturbation turns a given
singular continuous spectral measure into a pure point measure;
see \cite{djls} for many examples illustrating this phenomenon.

In \cite{dt}, the general criterion was applied to three prominent
models from one-dimensional quasicrystal theory, namely, the
Fibonacci model, the period doubling model, and the Thue-Morse
model. All these models can be generated by a substitution
process. This allows one to study the growth of transfer matrix
norms with the help of an associated dynamical system---the trace
map---and this provides in particular a very convenient way of
verifying the input to the general dynamical criterion.

In the present paper we will prove a more general version of the
dynamical result from \cite{dt}, involving also the weight
assigned by the spectral measure to the set of energies with
power-law bounded transfer matrices. This gives stronger dynamical
results in cases where such bounds hold for all energies in the
spectrum, for example, models with Sturmian potentials. We shall
also prove a stronger stability result. Namely, we will show that,
for a fixed energy, the power-law bound is stable with respect to
power-decaying perturbations. Here, the power-decay of the
perturbation that we can allow depends on the transfer matrix
power-law bound we start out with. Finally, we shall study a large
number of examples and derive dynamical results for them by
applying our main theorem, Theorem~\ref{main} below. The examples
discussed in this paper include, in particular, generalizations of
each of the three prominent substitution models studied in
\cite{dt}.

The organization of the paper is as follows. In Section~\ref{two}
we prove our main theorem which derives quantum dynamical lower
bounds from power-law bounds on transfer matrices.
Section~\ref{three} discusses the stability of such power-law
bounds on transfer matrices with respect to power-decaying
perturbations of the potential. Section~\ref{four} deals with a
class of models that are ``sparse'' in a certain sense and which
includes a variety of substitution models (in particular,
generalizations of Fibonacci, period doubling, and Thue-Morse),
the prime Schr\"odinger operator, and moderately sparse models
which were studied by Zlato\v{s} in \cite{z}. The hierarchical
model, which was studied in detail by Kunz et al.\ in \cite{KLS}
from a spectral point of view, will then be considered in
Section~\ref{five}. Finally, we present results for Sturmian
models (studied, e.g., in \cite{bist,dkl,irt}; see also the
reviews \cite{d2,s2}) in Section~\ref{six}.

%
%
%
%

\section{A Quantum Dynamical Lower Bound Derived From Power-Law Transfer Matrix Bounds}\label{two}

In this section, we prove a more general version of the main
result from \cite{dt}. The general idea of proof is the same and
the result derives lower bounds on the dynamical quantity
$\beta^-_{\delta_1}(p)$ from power-law bounds on transfer
matrices. However, the result established in this section gives
improved bounds in many cases, in particular in the case of
Sturmian potentials discussed later in the paper.

Recall the notion of a transfer matrix. Consider for some $E \in
\R$, a solution $\phi$ of the difference equation
\begin{equation}\label{eve}
\phi(n+1) + \phi(n-1) + V(n) \phi(n) = E \phi(n).
\end{equation}
Denote $\Phi(n) = ( \phi(n+1) , \phi(n) )^T$. The transfer matrix
$T(n,m;E)$ is defined by requiring
$$
\Phi(n) = T(n,m;E) \Phi(m)
$$
for every solution $\phi$ of \eqref{eve}. It is straightforward to
verify that for $n > m$,
$$
T(n,m;E) = T(V(n);E) \times \cdots \times T(V(m+1);E),
$$
where
$$
T(x;E) = \left( \begin{array}{cr} E - x & - 1 \\ 1 & 0 \end{array}
\right),
$$
and similarly for $n < m$.

With this notation at hand we can now state:

\begin{theorem}\label{main}

The following statements hold:\\[1mm]
{\rm (a)} Suppose that for some $K>0$, $C>0$, $\alpha >0$, the
following condition holds: For any $N>0$ large enough, there
exists a nonempty Borel set $A(N)\subset \R$ such that $A(N)
\subset [-K,K]$ and
\begin{equation}\label{maincond}
\|T(n,m;E)\| \le C N^{\alpha} \ \  \forall E \in A(N),\ \forall \
n,m: |n|\le N, |m| \le N
\end{equation}
{\rm (}resp., with $1 \le n \le N, \ 1 \le m \le N$ in the case of
$\ell^2(\N)${\rm )}. Let $N(T)=T^{1/(1+\alpha)}$ and let, for
$j=1,2$, $B_j(T)$ be the $j/T$-neighborhood of the set $A(N(T))$:
$$
B_j(T)=\{ E \in \R : \exists E' \in A(N(T)), |E-E'| \le j/T \}.
$$
Denote by $F(z)$ the Borel transform of the spectral measure of
the state $\psi=\delta_1$:
$$
F(E+i \varepsilon)=\int_{\R} \frac{d \mu (x)}{x-(E+i
\varepsilon)}.
$$

Then for the initial state $\psi=\delta_1$ and all $T
> 1$ large enough, the following bound holds:
\begin{equation}\label{fp1}
P(T) \equiv \sum_{n: |n| \ge N(T)} a(n,T) \ge {C \over T}
N^{1-2\alpha} (T) \int_{B_2(T)} dE (1+{\rm Im}^2
F(E+i\varepsilon)).
\end{equation}
In particular,
\begin{equation}\label{fp2}
P(T) \ge {C \over T} N^{1-2 \alpha} (T) (|B_1(T)|+\mu (B_1(T))),
\end{equation}
where $|B|$ denotes the Lebesgue measure.
This gives the following bound for the time-averaged moments:
\begin{equation}\label{fp9}
\langle |X|_{\delta_1}^p \rangle (T) \ge {C \over T} N^{p+1-2
\alpha} (T) (|B_1(T)|+\mu(B_1(T))).
\end{equation}
{\rm (b)} Suppose that there exists a set $A \subset [-K,K]$ of
positive measure $\mu(A)>0$ such that
$$
\| T(n,m;E)\| \le C (|n|^{\alpha} +|m|^{\alpha})
$$
for all $E \in A, \ n,m$. Then
\begin{equation}\label{fp7}
\beta_{\delta_1}^-(p) \ge \frac{p-3 \alpha}{1+\alpha}.
\end{equation}
{\rm (c)} Assume that
$$
\|T(n,m; E_0)\| \le C(E_0)(|n|^{\alpha}+|m|^{\alpha})
$$
for some $E_0$, uniformly in $n,m$, then
\begin{equation}\label{oneebound}
\langle |X|_{\delta_1}^p \rangle (T) \ge C T^{\frac{p-3
\alpha}{1+\alpha}} \left( T^{-1}+ \mu \left( \left[ E_0 - T^{-1} ,
E_0 + T^{-1} \right] \right) \right).
\end{equation}
Assume moreover that $E_0$ is an eigenvalue {\rm (}possible only
if $\alpha>1/2${\rm )}, so that there exists $\psi \in \ell^2, \
\psi \ne 0$ such that $H \psi=E_0 \psi$. Suppose that $\psi(1) \ne
0$ {\rm (}this is always true in the case of $\ell^2(\N)${\rm )}.
Then
\begin{equation}\label{fp10}
\beta_{\delta_1}^-(p)\ge {p+1-2\alpha \over 1+\alpha}.
\end{equation}
\end{theorem}

\begin{proof}
As in \cite{dt} we shall consider the case of $\ell^2(\Z)$,
because for $\ell^2(\N)$, the proof is similar but simpler. The
main part of the proof is virtually identical with that of
\cite{dt}. For the sake of completeness we shall briefly recall
the main lines.

The starting point is the Parseval equality:
$$
a(n,T) \equiv {1 \over T} \int_0^{\infty} e^{-2t/T} | \langle
\delta_n , \exp (-itH)\delta_1 \rangle |^2  \, dt =
\frac{\varepsilon}{2\pi} \int_{\R} | \langle \delta_n , R(E+i
\varepsilon)\delta_1 \rangle |^2 \, dE,
$$
where $R(z)=(H-zI)^{-1}$ and $\varepsilon = 1/T$.
For $z=E+i \varepsilon, \ \varepsilon>0$, we define $\phi=R(z)
\delta_1$,  $\Phi(n)=(\phi(n+1),\phi(n))^T$. For each $n>1$, one
has the inequality
\begin{equation}\label{fp3}
||\Phi (n)|| \ge ||T (n,1;z)||^{-1} ||\Phi(1)||
\end{equation}
and for each $n<0$,
\begin{equation}\label{fp4}
||\Phi(n)|| \ge ||T(n,0;z)||^{-1} ||\Phi (0)||.
\end{equation}
An upper bound for the norm of the transfer matrix with complex
$z$ is obtained using condition \eqref{maincond} and
\cite[Lemma~2.1]{dt}. Namely, let us fix some $T>1, \
\varepsilon=1/T$ and define $N \equiv N(T)= T^{1/(1+\alpha)}$.
Then for every $E \in B_2 (T)$ and $1 \le n \le N$
\begin{equation}\label{fp5}
||T(n,1;E+i \varepsilon)|| \le D N^{\alpha},
\end{equation}
where $D=C \exp (3C)$, and $C$ is the constant from
\eqref{maincond}. A similar bound holds for negative values of
$n$. Using the bounds \eqref{fp3}--\eqref{fp5}, one shows that for
every $E \in B_2(T)$,
\begin{equation}\label{fp11}
\sum_{n: |n| \ge N/2} | \langle \delta_n , R(E+i \varepsilon)
\delta_1 \rangle |^2 \ge c N^{1-2\alpha}
(|\phi(0)|^2+|\phi(1)|^2 +|\phi(2)|^2)
\end{equation}
with uniform constant $c>0$.
It was shown in \cite{dt} that under the conditions of the theorem
one always has
$$
|\phi(0)|+|\phi(1)|+|\phi(2)| \ge c>0
$$
with uniform constant. What one can also observe
(and this is a new point) is the fact that
$$
\phi(1)=\langle R(z) \delta_1 ,\delta_1 \rangle = F(z),
$$
where $F(z)$ is the Borel transform of the spectral measure
corresponding to the pair $(H,\delta_1)$. Therefore, it follows
from \eqref{fp11} that
$$
\sum_{n: |n| \ge N/2}| \langle \delta_n , R(E+i
\varepsilon)\delta_1 \rangle |^2 \ge c N^{1-2 \alpha} (1+{\rm
Im}^2 F(E+i \varepsilon)).
$$
Integrating this bound over $E \in B_2(T)$, one proves
\eqref{fp1}. Next, one observes that $1+{\rm Im}^2 F(z) \ge 2{\rm
Im} F(z)$. For any set $S$, denote by $S_\varepsilon$ the
$\varepsilon$-neighborhood of $S$. Following \cite{kkl}, one can
see that
\begin{align*}
\int_{S_\varepsilon} {\rm Im} F(E+i \varepsilon) dE & = \int_{\R}
d \mu (x) \int_{S_\varepsilon} {\varepsilon dE \over
(x-E)^2+\varepsilon^2}\\
& \ge \int_S d \mu (x) \int_{-\varepsilon}^{\varepsilon}
{\varepsilon du \over u^2+\varepsilon^2}\\
& = {\pi \over 2} \mu (S).
\end{align*}
Taking $S=B_1(T)$, we prove \eqref{fp2}. The bound \eqref{fp9}
immediately follows.

To prove part (b), one just takes $A(N)=A$ for every $N$. Since
$\mu (B_1(T)) \ge \mu (A(N(T)) =\mu (A)>0$, the result follows
from the bound \eqref{fp9}.

The bound \eqref{oneebound} of part (c) follows directly from
\eqref{fp9}, taking $A(N)=\{ E_0 \}$ for every $N$. Finally, to
prove the second part of (c), we go back to \eqref{fp1} to obtain
$$
\langle |X|_{\delta_1}^p \rangle (T) \ge {C \over T} N^{p+1-2
\alpha}(T) \int_{B_2(T)} {\rm Im}^2 F(E+i \varepsilon)
dE,
$$
where $B_2 (T)=[E_0-2 \varepsilon,E_0+2 \varepsilon]$. Under the
condition $\psi(1) \ne 0$, one has $\mu(\{ E_0 \})>0$. Thus,
$$
{\rm Im} F(E+i \varepsilon) \ge {c \varepsilon \over
(E-E_0)^2 +\varepsilon^2}.
$$
Integration over $B_2(T)$ yields \eqref{fp10}.
\end{proof}

\noindent\textit{Remark.} Part (b) of Theorem~\ref{main} remains
true if
\begin{equation}\label{asasa}
\| T(n,m; E) \| \le C(E) (|n|^{\alpha}+|m|^{\alpha})
\end{equation}
for all $n,m$ and $E \in A$ with $C(E) < \infty$ for $\mu$-almost
every $E$. To prove this, it is sufficient to take a smaller set
$A' \subset A$ of positive measure where $C(E) \le C < \infty$.
The bound \eqref{fp7} should be compared with the well-known
result of \cite{jl1,jl2}: If \eqref{asasa} holds for some $\alpha
\in [0, 1/2)$ on a set $A$ of positive $\mu$-measure, then the
restriction of $\mu$ to $A$ is $1-2\alpha$-continuous. In
particular,
$$
\beta_{\delta_1}^- (p) \ge p(1-2 \alpha).
$$
This bound is better than \eqref{fp7} for small $p$, but for $p$
large enough, \eqref{fp7} is always better. Moreover, \eqref{fp7}
holds also if $\alpha \ge 1/2$.

%
%
%
%

\section{Stability With Respect to Power-Decaying Perturbations}\label{three}

In this section, we discuss the stability of the crucial input to
our dynamical bounds, power-law bounds on transfer matrices, with
respect to perturbations of the potential. It is easy to see, and
was noted in \cite[Corollary~1.3]{dt}, that finitely supported
perturbations of the potential cannot destroy such a power-law
bound. Here we strengthen this to stability with respect to
power-decaying perturbations, where the allowed power depends on
the bound we can prove for the unperturbed problem.

\begin{theorem}\label{stability}
Assume that for some energy $E$ and some constant $C_1$, the
transfer matrices $T$ associated with $\Delta + V$ satisfy
\begin{equation}\label{v0transferbound}
\|T(n,m;E)\| \le C_1 |n-m|^\alpha \mbox{ for every } n,m \in \Z
\mbox{ with } nm \ge 0.
\end{equation}
Assume further that, for some $\varepsilon > 0$, the perturbation
$W$ satisfies
\begin{equation}\label{perturbbound}
|W(n)| \le C_2 (1 + |n|)^{-1 - 2 \alpha -\varepsilon} \mbox{ for
every } n \in \Z.
\end{equation}

Then the transfer matrices $T'$ associated with $\Delta + V + W$
satisfy
\begin{equation}\label{vtransferbound}
\|T'(n,m;E)\| \le C_3 |n-m|^\alpha \mbox{ for every } n,m \in \Z
\mbox{ with } nm \ge 0.
\end{equation}
\end{theorem}

\begin{proof}
We present the proof in the special case where we assume
\eqref{v0transferbound} only for $n \ge 0$ and $m = 0$ and then
prove \eqref{vtransferbound} for $n \ge 0$ and $m = 0$. A slight
variation of the argument below works for general $n,m \in \Z$
with $nm \ge 0$ (with a uniform constant $C_3$ in
\eqref{vtransferbound}).

Our strategy will be to work with solutions and employ a general
perturbation method developed by Kiselev et al.\ in \cite{krs}.

Consider the unperturbed equation \eqref{eve} and the perturbed
equation
\begin{equation}\label{evepert}
\psi(n+1) + \psi(n-1) + [V(n) + W(n)] \psi(n) = E \psi(n).
\end{equation}
Note that the transfer matrix $T'(n,0;E)$ is given by
$$
T'(n,0;E) = \left( \begin{array}{cc} \psi_{{\rm D}}(n+1) &
\psi_{{\rm N}}(n+1) \\ \psi_{{\rm D}}(n) & \psi_{{\rm N}}(n)
\end{array} \right),
$$
where $\psi_{{\rm D}} , \psi_{{\rm N}}$ solve \eqref{evepert} and
obey
$$
\left( \begin{array}{cc} \psi_{{\rm D}}(1) & \psi_{{\rm N}}(1)
\\ \psi_{{\rm D}}(0) & \psi_{{\rm N}}(0)
\end{array} \right) = I.
$$
Fix a complex reference solution $\phi$ of \eqref{eve}. For
example, we could set $\phi = \phi_{{\rm D}} + i \phi_{{\rm N}}$,
where $\phi_{{\rm D}}, \phi_{{\rm N}}$ solve \eqref{eve} and have
the same initial conditions as $\psi_{{\rm D}} , \psi_{{\rm N}}$.
By \eqref{v0transferbound} we have
\begin{equation}\label{phialpha}
|\phi(n)| \le C |n|^\alpha.
\end{equation}
Let $\psi$ be one of the basic solutions $\psi_{{\rm D}},
\psi_{{\rm N}}$ of \eqref{evepert}. Define $\rho(n)$ by
\begin{align*}
\left( \begin{array}{c} \psi(n) \\ \psi(n-1) \end{array} \right)
&= \frac{1}{2i} \left[ \rho(n) \left( \begin{array}{c} \phi(n) \\
\phi(n-1) \end{array} \right) - \overline{\rho(n)} \left(
\begin{array}{c} \overline{\phi(n)} \\ \overline{\phi(n-1)} \end{array}
\right)\right] \\
&= {\rm Im} \left[ \rho(n) \left( \begin{array}{c} \phi(n) \\
\phi(n-1) \end{array} \right) \right].
\end{align*}
Write $\phi(n)$ and $\rho(n)$ in polar coordinates,
$$
\phi(n) = |\phi(n)| e^{i \gamma(n)} \; \; \; \; \; \rho(n) = R(n)
e^{i \eta(n)},
$$
and define
$$
\theta(n) = \eta(n) + \gamma(n) \; \mbox{ and } \; U(n) = -
\frac{2 W(n)}{\omega} | \phi(n)|^2,
$$
where $i \omega$ is the Wronskian of $\overline{\phi}$ and $\phi$,
that is,
$$
2i {\rm Im} ( \phi(n+1) \overline{\phi(n)} ) = i \omega \mbox{ for
every } n.
$$
Clearly, the assertion of the theorem follows if we can show that
$R(n)$ remains bounded as $|n| \to \infty$. The key identity
(equation~(45) in \cite{krs}) is the following:
\begin{equation}\label{krsident}
R(n+1)^2 = R(n)^2 [ 1 + U(n) \sin ( 2 \theta(n)) + U(n)^2 \sin^2
(\theta(n)) ].
\end{equation}
It follows from \eqref{perturbbound} and \eqref{phialpha} that
$U(n)$ is summable. Thus, boundedness of $R(n)$ follows from this
and \eqref{krsident} (cf., e.g., \cite[Lemma~3.5]{kls2}). This
concludes the proof.
\end{proof}

The theorem above implies the stability of the number $\alpha$ and
of the sets $A(N)$, $B_1(T)$, $A$ under suitable power-decaying
perturbations of the potential. On the other hand, the measure of
the sets $\mu (B_1(T)), \ \mu (A)$ and the Borel transform $F(z)$
may change after such a perturbation. In particular, it is
possible that $\mu (A)=0$ for the perturbed operator in part (b)
of Theorem~\ref{main}. Thus, the bounds \eqref{fp7} and
\eqref{fp10} are in general not stable. Of course, we still get a
dynamical bound for the perturbed model. For example, we have the
following consequence of Theorem~\ref{main} and
Theorem~\ref{stability}:

\begin{coro}\label{pertdyn}
Assume that for some energy $E_0$ and some constant $C_1$, the
transfer matrices $T$ associated with $\Delta + V$ satisfy
$\|T(n,m;E_0)\| \le C_1 |n-m|^\alpha$ for every $n,m \in \Z$ with
$nm \ge 0$. Assume further that, for some $\varepsilon > 0$, the
perturbation $W$ satisfies $|W(n)| \le C_2 |n|^{-1 - 2 \alpha
-\varepsilon}$ for every $n \in \Z$. Then we have for the operator
$\Delta + V + W$,
$$
\beta_{\delta_1}^-(p) \ge \frac{p - 1 - 4 \alpha}{1 + \alpha}
$$
for every $p > 0$.
\end{coro}

\begin{proof}
By Theorem~\ref{stability}, we have that the transfer matrices
$T'$ associated with $\Delta + V + W$ satisfy $\|T'(n,m;E_0)\| \le C
|n-m|^\alpha$ for every $n,m \in \Z$ with $nm \ge 0$. Then, an
inspection of the proof of Theorem~\ref{main} shows that this
suffices to prove the bound \eqref{oneebound}
which  yields
$$
<|X|^p_{\delta_1} > (T) \ge C T^{{p-3 \alpha \over 1+\alpha} -1}
$$
and the assertion of the corollary follows.
 More precisely, one can
work independently on the two half-lines and hence needs bounds on
$\|T'(n,m;E)\|$ only for the case where $n,m$ have the same sign.
\end{proof}

%
%
%
%

\section{A Class of Pseudo-Sparse Potentials}\label{four}

In this section, we study a class of ``sparse'' potentials which
includes various substitution models and the prime model. These
potentials are not all sparse in the standard sense, but the point
is that the class we discuss contains sparse potentials, and also
a number of other potentials that have been considered before and
which can be studied within the same framework.

Let us consider the case where the potential $V$ is defined on the
half-line $\N$ and takes on two values $a,b \in \R$. We assume the
following for $n$ large enough, that is, for $n \ge N$:

\begin{itemize}
\item[(S1)] Occurrences of $b$ are always isolated, that is, if
$V(n) = b$ for some $n$, then $V(n-1) = V(n+1) = a$. \item[(S2)]
The value $a$ always occurs with odd multiplicity, that is, if
$V(n) = V(n+k+1) = b$ and $V(n+j) = a$, $1 \le j \le k$, then $k$
is odd.
\end{itemize}

Sparseness in this context refers to the $b$'s being isolated and
the results below holding for arbitrarily long gaps between
consecutive $b$'s. However, some of the concrete
applications---for example the applications to substitution
models---will not be sparse in a traditional sense.

We can prove the following:

\begin{theorem}\label{sparse2}
Suppose $V : \N \rightarrow \{a,b\} \subset \R$ is a potential
satisfying {\rm (S1)} and {\rm (S2)} above. We have for every $p
> 0$,
$$
\beta^-_{\delta_1}(p) \ge \frac{p-5}{2}.
$$
\end{theorem}

\begin{proof}
Up to an initial piece, the transfer matrices are given by
products of matrices of the following form:
$$
T(a,E)^{2l+1} \; \mbox{ and } \; T(b,E).
$$
Let $E_0 = a$. Then
$$
T(a,E_0)^{2l+1} = \left( T(a,E_0)^2 \right)^l T(a,E_0) = \left( -I \right)^l T(a,E_0) = \pm T(a,E_0).
$$
Up to sign, this gives rise to powers of
$$
T(a,E_0) T(b,E_0) = \left( \begin{array}{rr} 0 & -1 \\ 1 & 0
\end{array} \right) \left( \begin{array}{cr} a - b & -1 \\ 1 & 0
\end{array} \right) = \left( \begin{array}{cr} -1 & 0 \\ a - b &
-1 \end{array} \right).
$$
Clearly, such powers satisfy a bound which is linear in the number
of factors. Thus, the claim follows from \eqref{oneebound}.
\end{proof}

\noindent\textit{Remark.} We can apply Corollary~\ref{pertdyn} and
obtain that the dynamical bound in Theorem~\ref{sparse2} is stable
with respect to perturbations $W$ obeying $|W(n)| \le C_2 n^{- 3
-\varepsilon}$ for some fixed $\varepsilon > 0$ and every $n \in
\N$. Similarly, we have stability with respect to power-decaying
perturbations for all the dynamical bounds that will be shown in
this section and we will not make this explicit for each one of
them.

\medskip

Let us now discuss the case where the $a$'s occur with even
multiplicities. That is, we assume for $n$ large enough,

\begin{itemize}
\item[(S3)] The value $a$ always occurs with even multiplicity,
that is, if $V(n) = V(n+k+1) = b$ and $V(n+j) = a$, $1 \le j \le
k$, then $k$ is even.
\end{itemize}

In this case we can prove a dynamical bound even without assuming
the sparseness condition (S1). However, we need that $|a-b|$ is
not too large. Namely, we have the following result:

\begin{theorem}\label{sparse1}
Suppose $V : \N \rightarrow \{a,b\} \subset \R$ is a potential satisfying {\rm (S3)} above.\\[1mm]
{\rm (a)} If $|a-b| < 2$, then for every $p > 0$,
$$
\beta^-_{\delta_1}(p) \ge p-1.
$$
{\rm (b)} If $|a-b| = 2$, then for every $p > 0$,
$$
\beta^-_{\delta_1}(p) \ge \frac{p-5}{2}.
$$
\end{theorem}

\begin{proof}
The argument proceeds in a way similar to the proof above. Again,
up to an initial piece, the transfer matrices are given by
products of matrices of the following form:
$$
T(a,E)^{2l} \; \mbox{ and } \; T(b,E).
$$
Again, let $E_0 = a$. Then
$$
T(a,E_0) = \left( \begin{array}{rr} 0 & -1 \\ 1 & 0 \end{array} \right)
$$
and hence
$$
T(a,E_0)^{2l} = \left( T(a,E_0)^2 \right)^l = \left( -I \right)^l = \pm I.
$$
On the other hand, $T(b,E_0)$ is elliptic when $|a-b| < 2$ and
parabolic when $|a-b| = 2$. Thus, in the former case, products of
matrices of the form $T(a,E)^{2l}$ or $T(b,E)$ remain bounded,
while in the latter case such products satisfy a bound which is
linear in the number of factors. The claim thus follows from
\eqref{oneebound}.
\end{proof}

Let us note that a result like part (a) of Theorem~\ref{sparse1}
is implicitly contained in \cite{jss}, where mainly random polymer
models are studied.

It is clear that whole-line analogs of the above theorems hold. In
this case, we need (S1) and (S2) or (S3) to hold for $|n|$ large
enough.

More importantly, these results cover a variety of seemingly very
different cases: First consider the period doubling Hamiltonian,
which was already discussed in \cite{dt}. On the alphabet
$A=\{a,b\} \subseteq \R$, consider the period doubling
substitution $S(a)=ab$, $S(b)=aa$. Iterating on $a$, we obtain a
one-sided sequence
$$
u = abaaabababaaabaaab \ldots
$$
which is invariant under the substitution process. Define the
associated subshift $\Omega_{{\rm pd}}$ to be the set of all
sequences over $A$ which have all their finite subwords occurring
in $u$. Here, we can consider either one-sided or two-sided
sequences. This does not matter for the results in this paper, but
we remark that for substitution models, one generally considers
the two-sided case. For $\omega \in \Omega_{{\rm pd}}$, we define
the potential $V_\omega$ by $V_\omega (n) = \omega_n$. It is easy
to check that each $V_\omega$ satisfies (S1) and (S2) (even for
every $n \in \Z$) and hence an application of
Theorem~\ref{sparse2} allows us to recover \cite[Theorem~3]{dt}.
However, we can prove a more general result. Consider, for
example, substitutions of the form

\begin{equation}\label{subst1}
S(a) = a^{2k-1} b, \; S(b) = a^{2l}, \; k,l \ge 1.
\end{equation}
The case $k = 1, l = 1$ corresponds to the period doubling case.
The potentials generated by a substitution of the form
\eqref{subst1} (by generating a one-sided fixed point and passing
to the associated subshift, as in the period doubling case above)
are easily seen to obey (S1) and (S2). On the other hand,
substitutions of the form

\begin{equation}\label{subst2}
S(a) = a^{2k} b, \; S(b) = a^{2l}, \; k,l \ge 1
\end{equation}
give rise to potentials satisfying (S3) and hence
Theorem~\ref{sparse1} applies in these cases. Thus we may state
the following:

\begin{coro}\label{substcoro}
{\rm (a)} Let $S$ be a substitution of the form \eqref{subst1},
$\Omega$ the associated subshift, and for $\omega \in \Omega$, let
$V_\omega(n) = \omega_n$, $n \in \Z$. Then, for every $\omega \in
\Omega$, the potential $V_\omega$ gives rise to an operator
satisfying
$$
\beta^-_{\delta_1}(p) \ge \frac{p - 5}{2} \mbox{ for every } p >
0.
$$
{\rm (b)} Let $S$ be a substitution of the form \eqref{subst2},
$\Omega$ the associated subshift, and for $\omega \in \Omega$, let
$V_\omega(n) = \omega_n$, $n \in \Z$. Then, for every $\omega \in
\Omega$, the potential $V_\omega$ gives rise to an operator
satisfying
$$
\beta^-_{\delta_1}(p) \ge p - 1 \mbox{ for every } p > 0 \mbox{ if
} |a-b| < 2
$$
and
$$
\beta^-_{\delta_1}(p) \ge \frac{p - 5}{2} \mbox{ for every } p > 0
\mbox{ if } |a-b| = 2.
$$
\end{coro}

Consider the following class of substitutions:

\begin{equation}\label{gfs}
S(a) = a^m b^n , \; S(b) = a.
\end{equation}
The case $m = n = 1$ gives rise to the Fibonacci substitution.
Hence, the substitutions in \eqref{gfs} are usually called
generalized Fibonacci substitutions. If $n = 1$, the resulting
potentials are Sturmian and will be discussed in this more general
context in a later section. Here, we restrict our attention to the
case $n \ge 2$. These substitutions and the associated
Schr\"odinger operators were studied, for example, in
\cite{ka,sdr,t}.

If $n$ is even, it is easily seen that each $V_\omega$ satisfies
(S3) with the roles of $a$ and $b$ interchanged, that is, $b$'s
always occur with even multiplicity. Thus, we can derive a
dynamical bound for the associated operators by applying
Theorem~\ref{sparse1}.

If $n$ is odd, the model satisfies neither (S2) nor (S3) but we
can nevertheless employ a similar argument. As a warmup, let us
consider the case $n = 3$ (the special case $m = 1$, $n = 3$ is
usually called the nickel mean substitution). Then the transfer
matrices are given by products of matrices of the following form:
$$
T(a,E) \; \mbox{ and } \; T(b,E)^3.
$$
Let $E_0 = b+1$. Then
$$
T(b,E_0) = \left( \begin{array}{rr} 1 & -1 \\ 1 & 0 \end{array} \right)
$$
and hence
$$
T(b,E_0)^3 = - I.
$$
This would allow us to prove bounds on $\beta^-_{\delta_1}(p)$ in
the same way as in the proof of Theorem~\ref{sparse1}.

Let us now turn to the case of a general odd $n \ge 3$. Here we
can extend the above idea and prove a result which applies to the
substitutions in \eqref{gfs} with $n$ odd but which is much more
general. Denote

\begin{itemize}
\item[(S4)] There is some odd $k \ge 3$ such that the value $b$
always occurs with a multiplicity which is a multiple of $k$, that
is, if $V(n) = V(n+l+1) = a$ and $V(n+j) = b$, $1 \le j \le l$,
then $l = mk$ for some $m \in \N$,
\end{itemize}
Then we can prove the following:

\begin{theorem}\label{sparse3}
Suppose $V : \N \rightarrow \{a,b\} \subset \R$ is a potential
satisfying {\rm (S4)}. Then there is a set $\mathcal{E} \subset
\R$ of cardinality $k-1$ such that for every $E \in \mathcal{E}$, we have\\[1mm]
{\rm (a)} If $|a-E| < 2$, then for every $p > 0$,
$$
\beta^-_{\delta_1}(p) \ge p-1.
$$
{\rm (b)} If $|a-E| = 2$, then for every $p > 0$,
$$
\beta^-_{\delta_1}(p) \ge \frac{p-5}{2}.
$$
\end{theorem}

\begin{proof}
In this case, the transfer matrices are given by products of matrices of the following form:
$$
T(a,E) \; \mbox{ and } \; T(b,E)^k.
$$
It suffices to exhibit $k-1$ energies $E_0$ with
\begin{equation}\label{goode0}
T(b,E_0)^k = \pm I.
\end{equation}
This can be seen as follows: The matrix $T(b,E)^k$ is the
monodromy matrix of the constant potential $V(n) = b$, regarded as
a $k$-periodic potential. This gives rise to an operator with
$k-1$ gaps. However, since the operator with this potential has
spectrum $[b-2,b+2]$, all these gaps are degenerate. Every
degenerate gap corresponds to an energy where the monodromy matrix
is equal to $\pm I$, hence there are exactly $k-1$ energies $E_0$
for which we have \eqref{goode0}.
\end{proof}

Putting everything together, we obtain the following result for
the models generated by substitutions from \eqref{gfs}:

\begin{coro}\label{substcoro2}
Let $S$ be a substitution of the form \eqref{gfs}, $\Omega$ and the $V_\omega$'s as above.\\[1mm]
{\rm (a)} If $n \ge 2$ is even, then for every $\omega \in
\Omega$, the potential $V_\omega$ gives rise to an operator
satisfying
$$
\beta^-_{\delta_1}(p) \ge p - 1 \mbox{ for every } p > 0 \mbox{ if
} |a-b| < 2
$$
and
$$
\beta^-_{\delta_1}(p) \ge \frac{p - 5}{2} \mbox{ for every } p > 0
\mbox{ if } |a-b| = 2.
$$
{\rm (b)} If $n \ge 3$ odd, then $T(b,E_0)^n = \pm I$ has $n-1$
solutions $E_0 \in \R$ and for each such solution $E_0$, we have
that for every $\omega \in \Omega$, the potential $V_\omega$ gives
rise to an operator satisfying
$$
\beta^-_{\delta_1}(p) \ge p - 1 \mbox{ for every } p > 0 \mbox{ if
} |a-E_0| < 2
$$
and
$$
\beta^-_{\delta_1}(p) \ge \frac{p - 5}{2} \mbox{ for every } p > 0
\mbox{ if } |a-E_0| = 2.
$$
\end{coro}

The final substitution model we consider is the following:

\begin{equation}\label{gtms}
S(a) = a^m b^n , \; S(b) = b^n a^m.
\end{equation}
The case $m = n = 1$ gives rise to the Thue-Morse substitution.
Hence, the substitutions in \eqref{gtms} are usually called
generalized Thue-Morse substitutions. They were considered, for
example, in \cite{t}. If at least one of $m, n$ is even, (S3)
holds and we can apply Theorem~\ref{sparse1}. In the remaining
case, where both $m$ and $n$ are odd (and at least one  is $\ge
3$), (S4) holds and we can apply Theorem~\ref{sparse3}. Thus, for
models generated by generalized Thue-Morse substitutions, we
obtain the following dynamical bounds:

\begin{coro}\label{substcoro3}
Let $S$ be a substitution of the form \eqref{gtms}, $\Omega$ and the $V_\omega$'s as above. \\[1mm]
{\rm (a)} If at least one of $m,n$ is even, then for every $\omega
\in \Omega$, the potential $V_\omega$ gives rise to an operator
satisfying
$$
\beta^-_{\delta_1}(p) \ge p - 1 \mbox{ for every } p > 0 \mbox{ if
} |a-b| < 2
$$
and
$$
\beta^-_{\delta_1}(p) \ge \frac{p - 5}{2} \mbox{ for every } p > 0
\mbox{ if } |a-b| = 2.
$$
{\rm (b)} If we have $m \ge 3$ odd, then $T(b,E_0)^m = \pm I$ has
$m-1$ solutions $E_0 \in \R$ and for each such solution $E_0$, we
have that for every $\omega \in \Omega$, the potential $V_\omega$
gives rise to an operator satisfying
$$
\beta^-_{\delta_1}(p) \ge p - 1 \mbox{ for every } p > 0 \mbox{ if
} |b-E_0| < 2
$$
and
$$
\beta^-_{\delta_1}(p) \ge \frac{p - 5}{2} \mbox{ for every } p > 0
\mbox{ if } |b-E_0| = 2.
$$
An analogous result holds if we have $n \ge 3$ odd.\\[1mm]
{\rm (c)} If $m = n = 1$, then
$$
\beta^-_{\delta_1}(p) \ge p - 1 \mbox{ for every } p > 0.
$$
\end{coro}
Part (c) was proved in \cite{dt} and is stated for completeness.
One might expect the bound $\beta^-_{\delta_1}(p) \ge p - 1$ to
hold always. In fact, the paper \cite{t} claims, for every choice
of $m,n,a,b$, the existence of an energy, where the transfer
matrices remain bounded. However, the argument given in that paper
is incomplete and it would be interesting to prove or disprove
this claim.

Next, we consider the prime Schr\"odinger operator $H_{{\rm
prime}}$ on $\ell^2(\N)$ whose potential is given by
$$
V_{{\rm prime}}(n) = \left\{ \begin{array}{cl} a & \mbox{ if $n$
is not prime,} \\ b & \mbox{ if $n$ is prime.} \end{array} \right.
$$
This operator was studied, for example, in \cite{op,rkol}. Based
on numerics and heuristics contained in these two papers, one may
expect the following: On the one hand, for almost every energy
$E$, there is an $\ell^2$ solution to $H_{{\rm prime}} \phi = E
\phi$, that is, when one varies the boundary condition at the
origin, one gets pure point spectrum for almost every boundary
condition. On the other hand, the model displays non-trivial
transport for every boundary condition. We will confirm the latter
below (the proof discusses only the case of a Dirichlet boundary
condition, but it readily extends to every other boundary
condition). Let us briefly discuss the first point. It is natural
to view $V_{{\rm prime}}$ as a sparse potential. In fact, this
point of view was proposed in \cite{op}. However, the current
methods in the spectral analysis of models with sparse potentials
(see, in particular, \cite{kls2,r2}) are clearly insufficient to
conclude anything for the prime model. We regard this as an
interesting problem and refer the reader also to \cite{r3} for
further motivation to consider models of moderate sparseness.

Let us now turn to a dynamical result for the prime model.
Clearly, (S1) and (S2) are satisfied for $n$ large enough. Hence
we get:

\begin{coro}\label{primecoro}
For every $a,b \in \R$, the operator $H_{{\rm prime}}$ satisfies
$$
\beta^-_{\delta_1}(p) \ge \frac{p - 5}{2} \mbox{ for every } p >
0.
$$
\end{coro}

Finally, we discuss a model which is sparse in the standard sense.
Namely, pick some integer $\gamma \ge 2$ and define $n_k =
\gamma^k$ for $k \in \N$. Let $V_{{\rm sparse}}(n) = b$ if $n =
n_k$ for some $k$ and $V_{{\rm sparse}}(n) = a$ otherwise.
Schr\"odinger operators with potentials of this kind were studied
in \cite{z}. Clearly, when $\gamma$ is even, all $n_k$'s are even,
and when $\gamma$ is odd, all $n_k$'s are odd, so we have (S1) and
(S2). Thus, Theorem~\ref{sparse2} applies and we get

\begin{coro}\label{sparsecoro}
For every $a,b \in \R$ and $\gamma \in \N \setminus \{1\}$, the
potential $V_{{\rm sparse}}$ gives rise to an operator satisfying
$$
\beta^-_{\delta_1}(p) \ge \frac{p - 5}{2} \mbox{ for every } p >
0.
$$
\end{coro}
This can be improved if $\gamma \gg |a-b|$:

\begin{prop}\label{sparseprop}
Let
$$
\nu = \frac {2 \log \sqrt{2 + (a-b)^2}}{\log \gamma}.
$$
Then the potential $V_{{\rm sparse}}$ gives rise to an operator satisfying
$$
\beta^-_{\delta_1}(p) \ge \frac{p - 1 - 4 \nu}{1 + \nu} \mbox{ for
every } p > 0.
$$
\end{prop}

\begin{proof}
Write $C(a,b) = \sqrt{2 + (a-b)^2}$. Then
$$
\| T(a,E = a)^{2l + 1} T(b,E = a) \| = \left\| \left(
\begin{array}{cr} -1 & 0 \\ a - b & -1 \end{array} \right)
\right\| \le C(a,b).
$$
For $d_{n,m} = \# \{ m \le k \le n : V(k) = b\}$, we have $d_{n,m}
\le \log |n-m| / \log \gamma $ and hence
$$
\| T(n,m; E = a) \| \le C(a,b)^{d_{n,m}} \le C(a,b)^{\log |n-m| /
\log \gamma} = |n-m|^{\log C(a,b) / \log \gamma}.
$$
This yields the assertion.
\end{proof}

%
%
%
%

\section{A Hierarchical Model}\label{five}

The hierarchical model is defined through the potential

\begin{equation}\label{Vhier}
V(n) = \lambda f(\ord n),
\end{equation}
where $f$ is some real function and $\ord n$ is the number of
factors 2 in the prime decomposition of $n$. The sequence
\eqref{Vhier} has some nice symmetries. Because $\ord(-n)=\ord n$
for all $n$ and $\ord(l\cdot 2^m+k)=\ord k$ for $m\geq 1$, all $l$
and $|k|<2^m$, analogous identities hold for $V$. In particular,

\begin{equation}\label{Vsym}
V(l\cdot 2^m+k)=V(k)=V(-k)=V(l'\cdot 2^m-k)
\end{equation}
for any $l$ and $l'$, $m\geq 1$ and $|k|<2^m$. The Schr\"odinger
operator with such a potential appeared first in the works
\cite{LMR} and \cite{SWPZ} with the special choice
$$
f(m)=\sum_{k=0}^{m-1}R^k,
$$
where $R$ is a positive constant. The advantage of this choice is that in this case,
$$
x_m=\tr M_m(E)\equiv\tr T(2^m,0;E)
$$
satisfies an autonomous difference equation \cite{SWPZ},

\begin{equation}\label{recur}
x_{m+1}=x_m^2-2+Rx_m(x_m-x_{m-1}^2+2)\ ,\quad m\geq 1\ .
\end{equation}
The above recurrence and the symmetries \eqref{Vsym} made it
possible to obtain many rigorous results about the spectrum of the
corresponding Schr\"odinger operator. A detailed mathematical
study of this model was carried out by Kunz et al.\ \cite{KLS}.
Among other things, it was shown that for every $R>0$, the
spectrum is a Cantor set, and for $R\geq 1$, it is purely singular
continuous. From the point of view of the present article, it is
interesting that a countable infinite set of exceptional energies
in the spectrum could be identified explicitly. The $2^m$ zeros
$E_{mk}$, $1\leq k\leq 2^m$, of $x_m(E)$ are simple and $x_m=0$
implies $x_{m+1}=-2$ and $x_{m+l}=2$ for $l>1$;
compare~\eqref{recur}. From this it was possible to show that
$E_{mk}$, for $m\geq 0$ and $1\leq k\leq 2^m$, are lower (resp.,
upper) gap-edges in the spectrum of $H$ if $\lambda>0$ (resp.,
$\lambda<0$) and they are dense in the spectrum. For the
corresponding gap-edge states, the following result was obtained
(Proposition 15 in \cite{KLS}).

\begin{prop}\label{KLS}
Let $x_m(E)=0$ and let $\psi$ be a solution of $H\psi=E\psi$.
\begin{itemize}
\item[(i)]
If $\psi(0)=0$, then $\psi(k+2^{m+1})=-\psi(k)$ for every integer $k$.
\item[(ii)]
If $\psi(0)\neq 0$, then $\psi(2l\cdot 2^m)=(-1)^l\psi(0)$ and asymptotically, as $l\to\infty$,
\begin{equation}\label{psi}
\psi((2l+1)2^m)-\psi(2^m)\asymp (-1)^{l+1}\lambda_m\psi(0)f_R(l)
\end{equation}
where
\begin{equation}\label{fr}
f_R(l)=\left\{\begin{array}{ll}
            \frac{2}{2-R}l\ ,           &  R<2\\
            l\cdot\log_2 l\ ,         &  R=2\\
         \left(\frac{2}{R}\right)^{\varepsilon_l}\frac{R^2}{2(R-1)(R-2)}l^{\log_2 R}\ ,
         & R>2.
        \end{array}\right.
\end{equation}
Here $\lambda_m=\lambda R^m x_{m-1}(E)\cdots x_0(E)$,
$\varepsilon_l\in [0,1)$ is the fractional part of $\log_2 l$ and
$\asymp$ means equality in the leading order of $l$.
\end{itemize}
\end{prop}
We use this proposition to prove the following theorem.

\begin{theorem}\label{hierthm}
For every $\lambda\neq 0$ and $R>0$,
$$
\beta^-_{\delta_1}(p)\geq \frac{p-1-4\alpha}{1+\alpha},
$$
where
$$
\alpha=\alpha(R)=\max\{1,\log_2 R\}.
$$
\end{theorem}

\begin{proof}
We apply Proposition~\ref{KLS} with $m=0$ for which it provides
the precise asymptotic form of the solutions. Because $x_0(E)=E$,
these belong to $E=0$. Let $\psi_{{\rm D}}$ and $\psi_{{\rm N}}$
be the two solutions defined by the initial values
\begin{equation}\label{psi01}
\psi_{{\rm D}}(0)=\psi_{{\rm N}}(1)=0,\quad \psi_{{\rm
D}}(1)=\psi_{{\rm N}}(0)=1.
\end{equation}
According to part~(i) of Proposition~\ref{KLS}, $\psi_{{\rm D}}$
is a periodic solution with period 4, namely
\begin{equation}\label{psi0}
\psi_{{\rm D}}(2l)=0,\quad \psi_{{\rm D}}(2l+1)=(-1)^l.
\end{equation}
On the other hand,
\begin{equation}\label{psi1}
\psi_{{\rm N}}(2l)=(-1)^l,\quad \psi_{{\rm N}}(2l+1)\asymp
(-1)^{l+1}\lambda f_R(l).
\end{equation}
Equations \eqref{psi0} and \eqref{psi1} permit us to compute the
asymptotic form of $T(n,m;0)$. Because of $V(-n)=V(n)$, it
suffices to consider $n\geq m\geq 0$. In what follows, we use the
simplified notation $T(n,m)$. Let $\Psi^i(n)=(\psi^i(n+1)\
\psi^i(n))^T$ for $i=0,1$. Then $T(n,0)=(\Psi_{{\rm D}}(n)\
\Psi_{{\rm N}}(n))$. The determinant of any transfer matrix being
unity, the inverse is easy to compute. We find
\begin{align}T(n,m) & = T(n,0)T(m,0)^{-1}\\
\label{trans} & = \left(\begin{array}{ll}
                               \psi_{{\rm D}}(n+1) & \psi_{{\rm N}}(n+1)\\
                               \psi_{{\rm D}}(n)   & \psi_{{\rm N}}(n)
                               \end{array}\right)
\left(\begin{array}{rr}
      \psi_{{\rm N}}(m)   &-\psi_{{\rm N}}(m+1)\\
     -\psi_{{\rm D}}(m)    &\psi_{{\rm D}}(m+1)
      \end{array}\right).
\end{align}
With the short-hand notation
$$
F(l)=(-1)^l\psi_{{\rm N}}(2l+1),
$$
equations \eqref{psi0}, \eqref{psi1}, and \eqref{trans} then yield
\begin{align}
T(2l,2k)&=(-1)^{k+l}\left(\begin{array}{cc}
                         1 & F(l)\!-\!F(k)\\
                         0 & 1
                         \end{array}\right)\nonumber\\
T(2l+1,2k)&=(-1)^{k+l}\left(\begin{array}{cc}
                           0 & -1\\
                           1 & F(l)\!-\!F(k)
                           \end{array}\right)\nonumber\\
T(2l,2k+1)&=(-1)^{k+l+1}\left(\begin{array}{cr}
                             F(l)\!-\!F(k) & -1\\
                                1      & 0
                             \end{array}\right)\nonumber\\
T(2l+1,2k+1)&=(-1)^{k+l+1}\left(\begin{array}{cr}
                               -1      & 0\\
                               F(l)\!-\!F(k) & -1
                               \end{array}\right).
\end{align}
All these matrices have the same norm. Denoting the
Hilbert-Schmidt norm by $\|\cdot\|_2$, for $n=2l,2l+1$ and
$m=2k,2k+1$, we have
$$
\| T(n,m)\| \le \| T(n,m)\|_2 = \sqrt{2+[F(l)-F(k)]^2}
\asymp\sqrt{2+\lambda^2[f_R(l)-f_R(k)]^2}\ .
$$
Therefore,
$$
\| T(n,m;0)\| \leq 2\lambda f_R(n/2)
$$
for any $n$ large enough and $m\leq n$. If $R\neq 2$, the
assertion of the theorem obviously follows from the definition
\eqref{fr} of $f_R$ and Theorem~\ref{main}. If $R=2$, we note that
for any $\epsilon>0$,
$$
\| T(n,m;0)\| \leq \lambda n^{1+\epsilon}
$$
if $n$ is large enough. Therefore, by Theorem~\ref{main},
$$
\beta^-_{\delta_1}(p)\geq\frac{p-5-4\epsilon}{2+\epsilon}
$$
for any $\epsilon>0$ and, thus, for $\epsilon=0$ as well.
\end{proof}

\noindent\textit{Remark.} The proof shows that we can apply
Corollary~\ref{pertdyn} and obtain that the dynamical bound in
Theorem~\ref{hierthm} is stable with respect to perturbations $W$
obeying $|W(n)| \le C_2 |n|^{- 1 - 2\alpha -\varepsilon}$ for some
fixed $\varepsilon > 0$ and every $n \in \Z$.

\medskip

We note that instead of $m=0$, we could have used
Proposition~\ref{KLS} with any $m>0$ and any zero of $x_m(E)$.
This holds because of the following:

\begin{theorem}
For any $\lambda\neq 0$, $R>0$, $m\geq 0$, and
$k\in\{1,2,\ldots,2^m\}$, there exists a positive number
$C_{\lambda,R}(m,E_{mk})$ such that for any $n\geq n'\geq 0$,
$$
\| T(n,n';E_{mk})\|\leq C_{\lambda,R}(m,E_{mk})f_R(2^{-m-1}n)\ .
$$
\end{theorem}

\begin{proof}
We fix $m>0$ and a zero $E_{mk}$ of $x_m$. From equation
(\ref{trans}) it is clear that we have to bound the two particular
solutions (\ref{psi01}) of $H\psi=E_{mk}\psi$. According to
Proposition~\ref{KLS}, $\psi_{{\rm D}}$ is $2^{m+1}$-antiperiodic
and, thus, bounded. On the other hand,
\begin{equation}\label{psi1-emk}
\psi_{{\rm N}}(2l\cdot 2^m)=(-1)^l,\quad \psi_{{\rm
N}}((2l+1)2^m)-\psi_{{\rm N}}(2^m)\asymp (-1)^{l+1}\lambda_m
f_R(l).
\end{equation}
Thus, the task is to bound $\psi_{{\rm N}}(n)$ in the intervals
\begin{equation}\label{inter}
2l\cdot 2^m<n<(2l+1)2^m\quad{\rm and}\quad (2l+1)2^m<n<2(l+1)2^m\ .
\end{equation}
To proceed with the proof, let us recall equation (3.29) of \cite{KLS}, according
to which
$$
\psi_{{\rm D}}(2^m)=x_{m-1}\cdots x_0
$$
for any energy. Thus, $\psi_{{\rm D}}(2^m)\neq 0$ in the present
case ($E=E_{mk}$), for otherwise $x_i=0$ for some $i<m$ would
imply $|x_j|=2$ for every $j>i$, contradicting $x_m=0$. Then
$u_0:=\psi_{{\rm D}}/\psi_{{\rm D}}(2^m)$ is a solution of the
Schr\"odinger equation satisfying the boundary conditions
$u_0(0)=0$, $u_0(2^m)=1$ and, according to Proposition~\ref{KLS},
$u_0(k+2^{m+1})=-u_0(k)$ for any $k$. From the general theory of
second-order difference (differential) equations, it follows that
there exists a linearly independent solution $u_1$ with boundary
values $u_1(1)=1$, $u_1(2^m)=0$ and that we can write $\psi_{{\rm
N}}$ for $0\leq n\leq 2^m$ in the form
$$
\psi_{{\rm N}}(n)=\psi_{{\rm N}}(2^m)u_0(n)+\psi_{{\rm
N}}(0)u_1(n).
$$
Next, we observe that $u_1$ can be expressed in terms of $u_0$.
Indeed, from equation \eqref{Vsym} we can see that the sequence
$V(1),\ldots,V(2^m-1)$ is a palindrome,
$$
V(2^{m-1}-k)=V(2^{m-1}+k)\qquad k=1,\ldots,2^{m-1}-1
$$
and, hence,
$$
u_1(n)=u_0(2^m-n),\quad n=1,\ldots,2^m-1.
$$
Furthermore, the translational symmetry of the potential,
$$
(V(l\cdot 2^m+1),\ldots,V((l+1)2^m-1))=(V(1),\ldots,V(2^m-1)),
$$
valid for any $l$, implies that the translates of $u_0$ and $u_1$
can be used to give $\psi_{{\rm N}}$ in each of the intervals
\eqref{inter}. Altogether we find
$$
\psi_{{\rm N}}(n)=\psi_{{\rm N}}((2l+1)2^m)u_0(n-2l\cdot
2^m)+\psi_{{\rm N}}(2l\cdot 2^m)u_0((2l+1)2^m-n)
$$
if $2l\cdot 2^m\leq n\leq (2l+1)2^m$ and
$$
\psi_{{\rm N}}(n)=\psi_{{\rm
N}}(2(l+1)2^m)u_0(n-(2l+1)2^m)+\psi_{{\rm
N}}((2l+1)2^m)u_0(2(l+1)2^m-n)
$$
if $(2l+1)2^m\leq n\leq 2(l+1)2^m$. Together with
\eqref{psi1-emk}, in both intervals,
$$
|\psi_{{\rm N}}(n)|\leq \frac{\max |\psi_{{\rm D}}|}{|\psi_{{\rm
D}}(2^m)|}(|\psi_{{\rm N}}((2l+1)2^m)|+1).
$$
Since $l\leq n/2^{m+1}$, we obtain that for $n$ large enough
$$
|\psi_{{\rm N}}(n)|\leq  \frac{\max |\psi_{{\rm D}}|}{|\psi_{{\rm
D}}(2^m)|}(|\lambda_m| f_R(2^{-m-1}n)+|\psi_{{\rm N}}(2^m)|+1).
$$
Due to \eqref{trans}, the assertion of the theorem follows from
this bound.
\end{proof}

%
%
%
%

\section{Sturmian Potentials}\label{six}

In this section, we discuss dynamical bounds for the standard
one-dimensional quasicrystal model which is given by a
Schr\"odinger operator on the whole line whose potential is given
by

\begin{equation}\label{sturmpot}
V(n) = \lambda v_{\omega,\theta}(n), \; \mbox{ where } v_\theta(n) = \chi_{[1- \omega, 1)}
(n \omega + \theta \mod 1),
\end{equation}
where $\lambda \not=0$ is the coupling constant, $\omega \in (0,1)$ irrational is the
rotation number, and $\theta \in [0,1)$ arbitrary is the phase. For more information
on this family of operators, we refer the reader to the survey articles \cite{d2,s2}.

It is well known, and easy to see, that the spectrum of the operator
$H_{\lambda,\omega,\theta}$ with potential $V$ from \eqref{sturmpot} is independent
of $\theta$, that is, for every $\lambda, \omega$, there is a set $\Sigma_{\lambda,\omega}$
with $\sigma(H_{\lambda,\omega,\theta}) = \Sigma_{\lambda,\omega}$ for every $\theta$.

Consider the continued fraction expansion of $\omega$,
$$
\omega = \cfrac{1}{a_1+ \cfrac{1}{a_2+ \cfrac{1}{a_3 + \cdots}}}
$$
with uniquely determined $a_n \in \N$ (cf.~\cite{khin}). The
associated rational approximants $p_k/q_k$ are defined by

\begin{alignat*}{3}
p_0 &= 0, &\quad    p_1 &= 1,   &\quad  p_k &= a_k p_{k-1} + p_{k-2},\\
q_0 &= 1, &     q_1 &= a_1, &       q_k &= a_k q_{k-1} + q_{k-2}.
\end{alignat*}
The number $\omega$ is said to have bounded density if

\begin{equation}\label{domega}
d(\omega) = \limsup_{n \rightarrow \infty} \frac{1}{n} \sum_{k=1}^n a_k < \infty.
\end{equation}
The set of bounded density numbers is uncountable but has Lebesgue measure zero.
The following was shown in \cite{dl2} (see also \cite{irt} for the case of zero phase):

\begin{theorem}\label{pl2_2}
Suppose $\omega$ is a bounded density number. For every $\lambda$, there is a constant
$C$ such that for every $\theta$, every $E \in \Sigma_{\lambda,\omega}$, and every
$n,m \in \Z$, we have

\begin{equation}
\| T_{\lambda,\omega,\theta}(n,m;E) \| \le C |n-m|^{\alpha (\lambda, \omega)},
\end{equation}
with

\begin{equation}\label{alphagen}
\alpha (\lambda, \omega) = D \cdot d(\omega) \cdot \log C_\lambda,
\end{equation}
where $D$ is some universal constant, $C_\lambda$ is given by

\begin{equation}\label{clambda}
C_\lambda = 2 + \sqrt{8 + \lambda^2},
\end{equation}
and $d(\omega)$ is as in \eqref{domega}.
\end{theorem}

This yields the following:

\begin{coro}\label{sturmtheo1}
Let $\omega$ be a bounded density number. Then, for every $\lambda, \theta$, the
operator $H_{\lambda, \omega, \theta}$ satisfies
$$
\beta^-_{\delta_1}(p) \ge \frac{p - 3 \alpha(\lambda, \omega)}{ 1
+ \alpha(\lambda, \omega)} \mbox{ for every } p > 0,
$$
with $\alpha(\lambda, \omega)$ given by \eqref{alphagen}.
\end{coro}

Since $\mu (\Sigma_{\lambda,\omega})=1$, this is an immediate
consequence of \eqref{fp7}. This bound is better than the
corresponding result in \cite{dt} (which follows from \eqref{fp9},
bounding from below $|B_1(T)|$). One should stress that as opposed
to all the other examples discussed earlier, the dynamical bound
in Corollary~\ref{sturmtheo1} is not stable with respect to
perturbations of the potential. This is due to the fact that $\mu
(\Sigma_{\lambda,\omega})$ may vanish for the perturbed measure.
However, by Corollary~\ref{pertdyn}, we have the following result:

\begin{coro}\label{sturmtheo2}
Let $\omega$ be a bounded density number and let $\lambda$ be
arbitrary. If $\alpha(\lambda, \omega)$ is given by
\eqref{alphagen} and $W$ satisfies
$$
|W(n)| \le C_2 (1 + |n|)^{-1 - 2 \alpha(\lambda,\omega)
-\varepsilon} \mbox{ for every } n \in \Z
$$
for some $\varepsilon > 0$,  then, for every $\theta$, the
operator $H_{\lambda, \omega, \theta} + W$ satisfies
$$
\beta^-_{\delta_1}(p) \ge \frac{p - 1 - 4 \alpha(\lambda,
\omega)}{ 1 + \alpha(\lambda, \omega)} \mbox{ for every } p > 0.
$$
\end{coro}

As in the case
$\omega = (\sqrt{5} - 1)/2$ and $\theta = 0$, studied in
\cite{dt}, it is possible to improve this lower bound
somewhat by exhibiting a suitable set $A(N)$ (stable under perturbation),
studying its
Lebesgue measure, and applying \eqref{fp9}. The set $A(N)$ will
again be given by the spectra of suitable periodic approximants,
and the Lebesgue measure can again be bounded through a fine
analysis of the trace map, akin to what is done in
\cite{dt,kkl,r}; compare also \cite{lw}. We leave the details to
the interested reader.

\end{document}